\documentclass[useAMS,usenatbib]{mn2e}

\usepackage{amsmath}
\usepackage{graphicx}
\usepackage{color}
\usepackage[colorlinks,citecolor=blue]{hyperref}

\title[Constraints on neutrino mass from Cosmic Microwave Background and Large Scale Structure]
{Constraints on neutrino mass from Cosmic Microwave Background and Large Scale Structure}
\author[Z. Pan and L. Knox]{Z.~Pan, $^1$\thanks{Email: zhpan@ucdavis.edu}
        L.~Knox$^1$\thanks{Email: lknox@ucdavis.edu}\\
$^1$Department of Physics, University of California, One Shields Avenue, Davis, CA, USA 95616}

\date{\today}

\def\be{\begin{equation}}
\def\ee{\end{equation}}
\def\bea{\begin{eqnarray}}
\def\eea{\end{eqnarray}}

\def \nn{\nonumber}
\def \eq{equation}

\def\nnu{N_{\rm eff}}
\def \Planck {{\it Planck}\ }

 \def\C{{\mathcal C}_\ell}
 \def\N{N_\ell}

\begin{document}
\label{firstpage}

\maketitle

\begin{abstract}
Our tightest upper limit on the sum of neutrino mass eigenvalues $M_\nu$ comes from cosmological
observations that will improve substantially in the near future, enabling
a detection. The combination of the Baryon Acoustic Oscillation feature measured from the Dark Energy 
Spectroscopic Instrument  and a Stage-IV  Cosmic Microwave Background  experiment
has been forecasted to achieve $\sigma(M_\nu) < 1/3$ of the lower limit on $M_\nu$ from atmospheric and 
solar neutrino oscillations \citep{2013arXiv1309.5383A,2012PhRvD..86a3012F}. Here we examine in detail the physical effects of neutrino mass on 
cosmological observables that make these constraints possible. We also consider how these constraints 
would be improved to ensure at least a $5\sigma$ detection.
\end{abstract}


\bigskip\bigskip

\section{Introduction}
Basic questions about the neutrino mass matrix remain unanswered, such as whether the CP-violating phase is non-zero, whether the neutrinos are Majorana or Dirac, and whether the hierarchy of masses is normal or inverted \citep{2012PhRvD..86a3012F, 2012PhRvD..86g3012F, 2002PhRvD..65g3023B}. Significant experimental  and observational efforts are underway and being planned to answer these questions. Doing so may shed light on possible extensions beyond the standard model of particle physics.

The question of the type of mass hierarchy may be settled by cosmological observations. The best lower limit on the sum of the mass eigenstate masses, $M_\nu$, comes from analysis of solar and atmospheric neutrino oscillation data \citep{2014PhRvD..90i3006F}. If cosmological determinations of this quantity tighten up near this lower bound, then the inverted hierarchy will be ruled out.

Current data lead to $0.058\ {\rm eV} \la M_\nu \la 0.21\ {\rm eV}$, where the upper limit comes from cosmic microwave background (CMB) and baryon acoustic oscillation (BAO) data \citep{PlanckCollaborationXIII.2015}. Bringing this upper limit down is a major science goal of a Stage-IV (S4) cosmic microwave background project, CMB-S4, and also of the galaxy survey project Dark Energy Spectroscopic Instrument (DESI). These projects are
forecasted to determine $M_\nu$ with a one-standard deviation of $45$ meV (CMB-S4 alone) and $16$ meV (CMB-S4 combined with DESI BAO)  \citep{2013arXiv1309.5383A}. These uncertainties are small enough to guarantee a detection of $M_\nu \ne 0$ at a 3$\sigma$ or greater confidence level.

Of course any conclusions from such cosmological data will be model dependent. How convincing will these data be that we are indeed seeing the impact of neutrino mass, and not misinterpreting some other signals? The forecasted precision is also not quite as strong as one would like; is there a way to guarantee at least a 5$\sigma$ detection of $M_\nu \ne 0$? Here we address these questions.
To address the first question we examine the particular signatures of neutrino mass that lead to the above forecasts. To address the second we look at what additional types of data can further tighten the expected uncertainties.

The paper is organized as follows.
In Section 2, we briefly introduce the cosmological signatures of massive neutrinos.
In Section 3,  we focus on changes in the cosmic expansion rate and structure growth rate due to massive neutrinos. In Section 4, we analyze the influence of massive neutrinos on the CMB lensing potential power spectrum. Forecasts on the constraints of total neutrino mass from CMB and Large Scale Structure (LSS) measurements are given in Section 5 and conclusions are presented in Section 6.

\section{Signatures of non-zero neutrino mass}
\label{sec:sign}
The cosmological signatures of massive neutrino have been investigated since decades ago, e.g. \citep{Doroshkevich1980, Doroshkevich1981, Doroshkevich1981b}.
We can more broadly view the cosmological neutrino program as a study of the “dark radiation” that we know exists as a thermal relic of the big bang.
By dark radiation here we mean anything, other than photons, thermally produced in the early universe that is relativistic at least through decoupling. We know that such a background of nearly massless non-photon radiation exists with high confidence from light element abundances and the cosmic microwave background damping tail. Both are sensitive to the history of the expansion rate, which depends on the mean density via the Friedman equation. Combining Helium abundance and CMB data constrains the effective number of relativistic species to be $\nnu = 2.99\pm0.39$ \citep{PlanckCollaborationXIII.2015}.

Is this background entirely that of the 3 active neutrino species? Is any part of it from something else? Could there be a significant excess of neutrinos over anti-neutrinos? These are interesting questions, also to be addressed by future CMB observations that will significantly tighten up constraints on $\nnu$. Our confidence that the dark radiation is indeed that of cosmological neutrinos with phase-space distributions as expected from the standard thermal history will be greatly increased if the constraints on $\nnu$ tighten up to $\sigma(\nnu) = 0.02$, as forecasted, consistent with the expected value of 3.046. For the purposes of this paper, we will assume this is what will happen.

If we assume that the dark radiation background is entirely the active neutrinos with the expected phase space distributions, the assumption of non-zero neutrino mass leads to very specific predictions for cosmological observables. First we consider the expansion rate as a function of redshift. The rest-mass energy of the neutrinos begins to slow down the decline of energy density with expansion as they become non-relativistic, leading to an increase in $H(z)$ relative to the $M_\nu = 0$ expectation. This increase would persist to $z=0$ if we were holding the other contents of the low-redshift universe constant. However, for our purposes of exploring observable consequences of $M_\nu \ne 0$ it makes much more sense to hold the angular size of the sound horizon on the last-scattering surface constant, since this quantity is so well-determined from CMB observations \citep{PlanckCollaborationXIII.2015, Hinshaw2013}. To do so one must decrease the density of dark energy. Assuming the dark energy is a cosmological constant, the shape of $\Delta H(z)$ has a very particular form, as shown in Fig.~\ref{fig:DzHz}, changing sign very near $z = 1$, with the onset of dark energy domination.

Were we able to trace out this departure of $H(z)$ from the $M_\nu = 0$ shape, it would contribute to our confidence we are seeing the impact of non-zero neutrino mass. However, as we will see, the DESI determinations of $H(z)$ will be insufficient to resolve this very small signal across redshift. That is not to say the signal is altogether observably invisible. These changes to $H(z)$ affect comoving angular diameter distance $D_A(z)$ in ways that are detectable by DESI. It is just that it will be difficult, if not impossible, to make the case that there is the sign change in the $H(z)$ correction near $z \simeq 1$. The changes to $H(z)$ also directly impact the growth of structure, with observable consequences for the redshift-space distortion (RSD) and the CMB lensing potential power spectrum, which we will discuss in Section \ref{sec:galaxy} and \ref{sec:lensCl}  respectively.

\section{Influence of massive neutrinos on galaxy survey observables}
\label{sec:galaxy}

To quantify the influence of massive neutrinos, we compare a fiducial cosmology with massless neutrinos and a cosmology with massive neutrinos. The fiducial cosmology is a flat $\Lambda$CDM universe with the \Planck best fit parameters \citep{PlanckCollaborationXVI.2013},  i.e. $\omega_b = 0.022032, \omega_{\rm m} = 0.14305, A_s = 2.215\times10^{-9}, n_s= 0.9619,\tau = 0.0925, H_0 = 67.04 \ {\rm km/s/Mpc}, M_\nu=0$ meV.
The set of $M_\nu\neq0$ cosmologies have parameters $\theta = (\Theta, M_\nu)$, where $\Theta$ are $\Lambda$CDM parameters.
Given a specific $M_\nu$, we choose $\Theta$ by minimizing $\chi^2(\Theta, M_\nu)$,  with
\bea
\chi^2(\Theta, M_\nu)
&=&F_{\alpha\beta}\lambda_\alpha\lambda_\beta\nn\\
&=&F_{ij}\lambda_i\lambda_j + 2 F_{i\nu}\lambda_i M_\nu + F_{\nu\nu}M_\nu^2,
\eea
where $F$ is the Fisher matrix for the CMB observations, $\lambda_i = (\Theta-\Theta_{\rm fid})_i$  with $i$ indexing the 6 $\Lambda$CDM parameters and summation over repeated indexes $\alpha, \beta, i, j$ is implied. Minimizing $\chi^2(\Theta, M_\nu)$ requires
\be
0=\partial\chi^2/\partial \lambda_i \rightarrow \lambda_i = - (G^{-1})_{ij} F_{j\nu} M_\nu,
\ee
where $G$ is a subset of the Fisher matrix $F$, $G_{ij} \equiv F_{ij}$.

\begin{figure}
\includegraphics[scale=0.4]{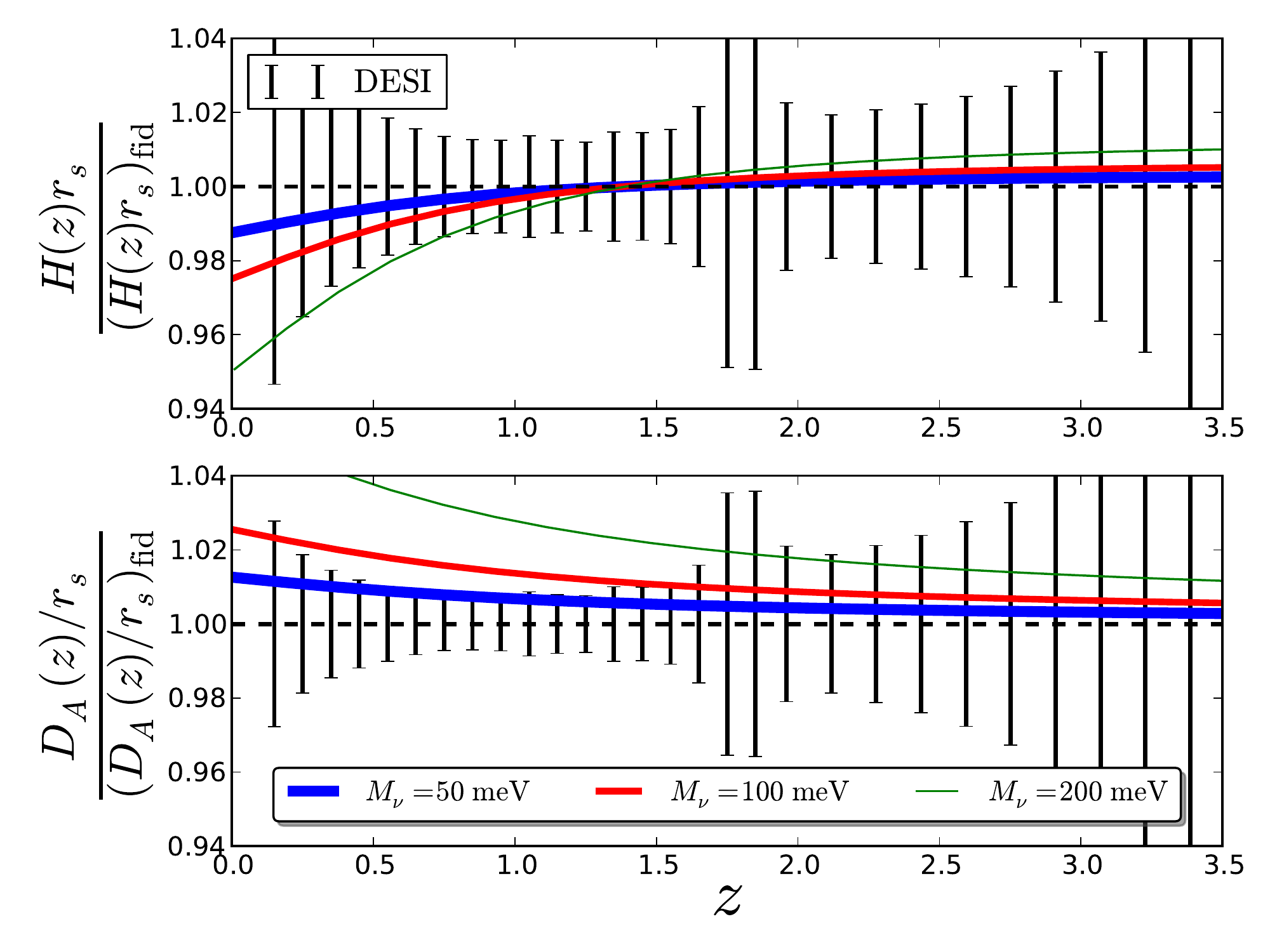}
\caption{The dependence of expansion rate $H(z)$ and comoving angular diameter distance $D_A(z)$ on $M_\nu$,
where we minimize the $\chi^2(\Theta, M_\nu)$ by adjust the $6$ $\Lambda$CDM parameters $\Theta$ when increasing $M_\nu$ from $0$ to $50, 100, 200$ meV. }
\label{fig:DzHz}
\end{figure}

In Fig. \ref{fig:DzHz}, we show the influence of $M_\nu= 50, 100$ and $200$ meV on 
expansion rate $H(z)$ and comoving angular diameter distance $D_A(z)$. 
We see that $H(z\la 1 )$ decreases and $H(z\ga1)$ increases compared to the fiducial cosmology, and the comving distance $D_A(z)$ increases accordingly. Though the departure of $H(z)$ from the $M_\nu = 0$ shape is undetectable by DESI BAO, the changes in $D_A(z)$ are readily detectable \citep{2014JCAP...05..023F}. \\

One of the observable consequences of these changes to $H(z)$ is the impact on the structure growth rate.
How one describes this impact depends on what one is using
for a comparison model. We use, as a comparison model, a cosmology with massless neutrinos in place of the massive ones.
One could also use as a comparison model one with additional cold dark matter in place of the neutrinos. We
use the former, consistent with our underlying assumption that we have 3 neutrino species with phase-space distributions as expected
from the standard thermal history. 

\begin{figure}
\label{fig:growth}
\includegraphics[scale=0.4]{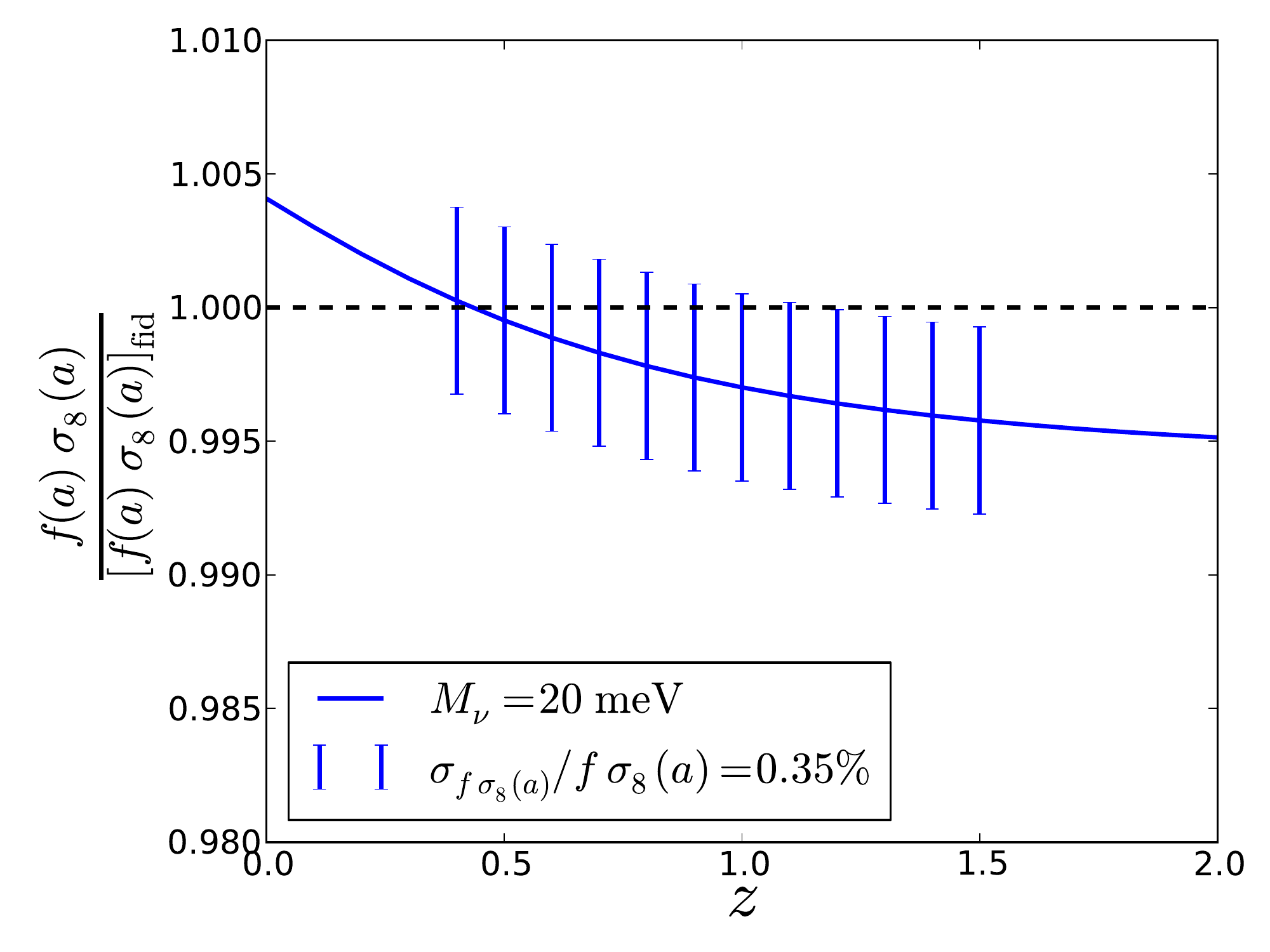}
\caption{The dependence of structure growth rate on $M_\nu$,
where we minimize the $\chi^2(\Theta, M_\nu)$ by adjust the $6$ $\Lambda$CDM parameters $\Theta$ when increasing $M_\nu$ from $0$ to $20$ meV.}
\label{fig:growth}
\end{figure}

The structure growth rate that is usually quantified by $d \ln \sigma_8(a)/d\ln a$, can be determined by RSD from galaxy surveys, where $\sigma_8(a)$ is the amplitude of mass fluctuations $\sigma_R$ on scale of $8h^{-1}$ Mpc, i.e., $\sigma_8(a)\equiv\sigma(R=8h^{-1} {\rm Mpc}; a)$.
Here
\be
\sigma_R^2(a) \equiv \int_0^\infty \frac{k^3}{2\pi^2} P_\delta(k;a) W^2(kR) d\ln k,
\ee
where $P_\delta(k,a)$ is the matter power spectrum defined by, 
$\left\langle \delta(\mathbf{k};a)\delta^\star(\mathbf{k'};a)\right\rangle
  =  P_\delta(k;a)\delta^{(3)}(\mathbf{k}-\mathbf{k'})$,
and $W(kR) = 3j_1(kR)/kR$ is the window function. 
Introducing the growth function $D(a)$, 
\be
D(a) \equiv\frac{ \delta(a)}{\delta(a=1) } = \frac{\sigma_8(a)}{\sigma_8(a=1)},
\ee
where $\delta(a)$ is the matter overdensity at redshift $z = 1/a-1$,
and defining perturbation growth rate $f(a) \equiv d\ln D(a)/d\ln a$ \citep{2001MNRAS.325...77P}, we may rewrite the structure growth rate as $d \ln \sigma_8(a)/d\ln a = f(a) \sigma_8(a)$.

In the above definition, the growth function $D(a)$ and the structure growth rate $f(a) \sigma_8(a)$ depend on $\sigma_8(a=1)$ which varies for our fiducial $M_\nu = 0$ model and the model with massive neutrinos. For easier comparison, we introduce the growth function $D_e(a)$ normalized at early time, say at $a_e =1/1100$ when massive neutrinos are relativistic, 
\be
D_e(a) \equiv\frac{ \delta(a)}{\delta(a_e) } = \frac{\sigma_8(a)}{\sigma_8(a_e)},
\ee
and the structure growth rate can be rewritten as
\be
f(a)\sigma_8(a) = \frac{d D_e(a)}{d \ln a} \sigma_8(a_e),
\ee
where $\sigma_8(a_e)$ is the same for the model with massive neutrinos and the fiducial model.
So the variation of $f(a)\sigma_8(a)$ only depends on the growth rate $d D_e(a)/d a$.

In Fig. \ref{fig:growth}, we show the impact of massive neutrinos with mass $20$ meV on the structure growth rate. The structure growth rate is decreased at high redshift because of  the enhanced expansion rate (Fig. \ref{fig:DzHz}). Note though that the transition of structure growth rate from suppressed to enhanced is delayed to $z\simeq 0.5$ compared to $z\simeq 1$ when the expansion rate transitions. The reason for this delay is that the growth rate $d D_e(a)/ da$ is determined by two factors: the expansion rate and gravitational attraction. The slower growth at $z \la 1$ leads to weaker gravitational potentials. Around $z\simeq 1$, the expansion rate is the same for the two models, but the gravitational potential is weaker for the $M_\nu \ne 0$ model. Therefore the growth rate remains suppressed,until some later time $z\simeq 0.5$ when the weaker gravitational potential is compensated by even slower expansion. 

DESI will provide a comprehensive survey of spectroscopic galaxies and quasars covering redshifts $0.1<z<3.5$, with precision in each redshift bin $\Delta z=0.1$ better than $\sigma_{f\sigma_8(a)}/f\sigma_8(a) = 0.35\%$ from $0.4 < z <1.5$ \citep{2013arXiv1309.5385H}. In Fig. \ref{fig:growth}, we show the influence of $M_\nu= 20$ meV on the structure growth rate $f(a) \sigma_8(a)$, which is detectable by DESI RSD.

\section{Influence of massive neutrinos on the CMB lensing power spectrum}
\label{sec:lensCl}

\subsection{Introduction to the lensing power spectrum}
We begin with a brief review of gravitational lensing of the CMB. 
For details see e.g.~the review by \citet{Lewis2006a}.
Gradients in the gravitational potential, $\Phi$, distort the trajectories
of photons traveling to us from the last scattering surface.
The deflection angles, in Born approximation, are $\mathbf{d} = \nabla\phi$,
where the lensing potential, $\phi$, is a weighted radial projection of $\Phi$.
The key quantity for calculating the impact of lensing on the temperature
power spectrum is the angular power spectrum of the projected potential,
$C_\ell^{\phi\phi}$, which we also call the lensing power spectrum.
Taking advantage of the Limber approximation \citep{Limber}, it can be
written as a radial integral over the three dimensional gravitational
potential power spectrum $P_\Phi$
\begin{\eq}
  \ell^4 C_\ell^{\phi\phi} \simeq 4\int_0^{\chi_\star}
  \mathrm{d}\chi\ (k^4P_\Phi)\left(\frac{\ell}{\chi};a\right)
  \left[1-\frac{\chi}{\chi_\star}\right]^2 ,
\label{eqn:lensCl}
\end{\eq}
where $\chi$ is the comoving distance from the observer, $a=a(\chi)$,
a $\star$ subscript indicates the last scattering surface,
$(1-\chi/\chi_\star)^2$ is the lensing kernel, and the power
spectrum $P_\Phi$ is defined as
\begin{\eq}
  \left\langle \Phi(\mathbf{k};a)\Phi^\star(\mathbf{k'};a)\right\rangle
  =  P_\Phi(k;a)\delta^{(3)}(\mathbf{k}-\mathbf{k'}).
\end{\eq}
To calculate $P_\Phi$ we assume a power-law primordial power spectrum
$P^p_\Phi(k)$, 
\begin{\eq}
  \frac{k^3}{2\pi^2} P^p_\Phi(k) = A_s\left(\frac{k}{k_0}\right)^{n_s-1},
\end{\eq}
where $k_0$ is an arbitrary pivot point, $A_s$ and $n_s$ are the primordial
amplitude and power law index respectively.

The gravitational potential at late times, $\Phi(k,a)$, is related to the
primordial potential $\Phi^p(k)$ by \citep[e.g.][]{KodSas84}
\begin{\eq}
  \Phi(k,a) = \frac{9}{10}\Phi^p(k)T(k) s(k; a) g(a) ,
\end{\eq}
where the potential on very large scales is suppressed by a factor $9/10$
through the transition from radiation domination to matter domination.
For modes which enter the horizon during radiation domination when the
dominant component has significant pressure ($p\approx \rho/3$) the amplitude
of perturbations cannot grow and the expansion of the Universe forces the
potentials to decay.  In a cosmology with massless neutrinos, 
for modes which enter the horizon after matter-radiation
equality (but before dark energy domination) the potentials remain constant.
The transfer function $T(k)$ takes this into account, being unity for very
large scale modes and falling approximately as $k^{-2}$ for small scales.
The transfer function $T(k)$ is independent of scale factor $a$ 
because in a cosmology with massless neutrinos, potentials on all scales keep constant in matter domination.  
The impact of massive neutrinos on the 
potentials is described by the function, $s(k;a)$ , which is unity on scales
above the free-streaming scale and decreases with time on scales below.
Once the cosmological constant starts to become important the potentials on
all scales begin to decay. This effect is captured by the growth function,
$g(a)$, which is unity during matter domination. With these definitions we have
\begin{\eq}
  (k^4P_\Phi)(k;a)=\frac{81}{50}\pi^2
  g^2(a)\ s^2(k; a)\ k T^2(k)\  A_s \left(\frac{k}{k_0}\right)^{n_s-1} .
\label{eqn:Pphi}
\end{\eq}

\begin{figure}
\includegraphics[scale=0.42]{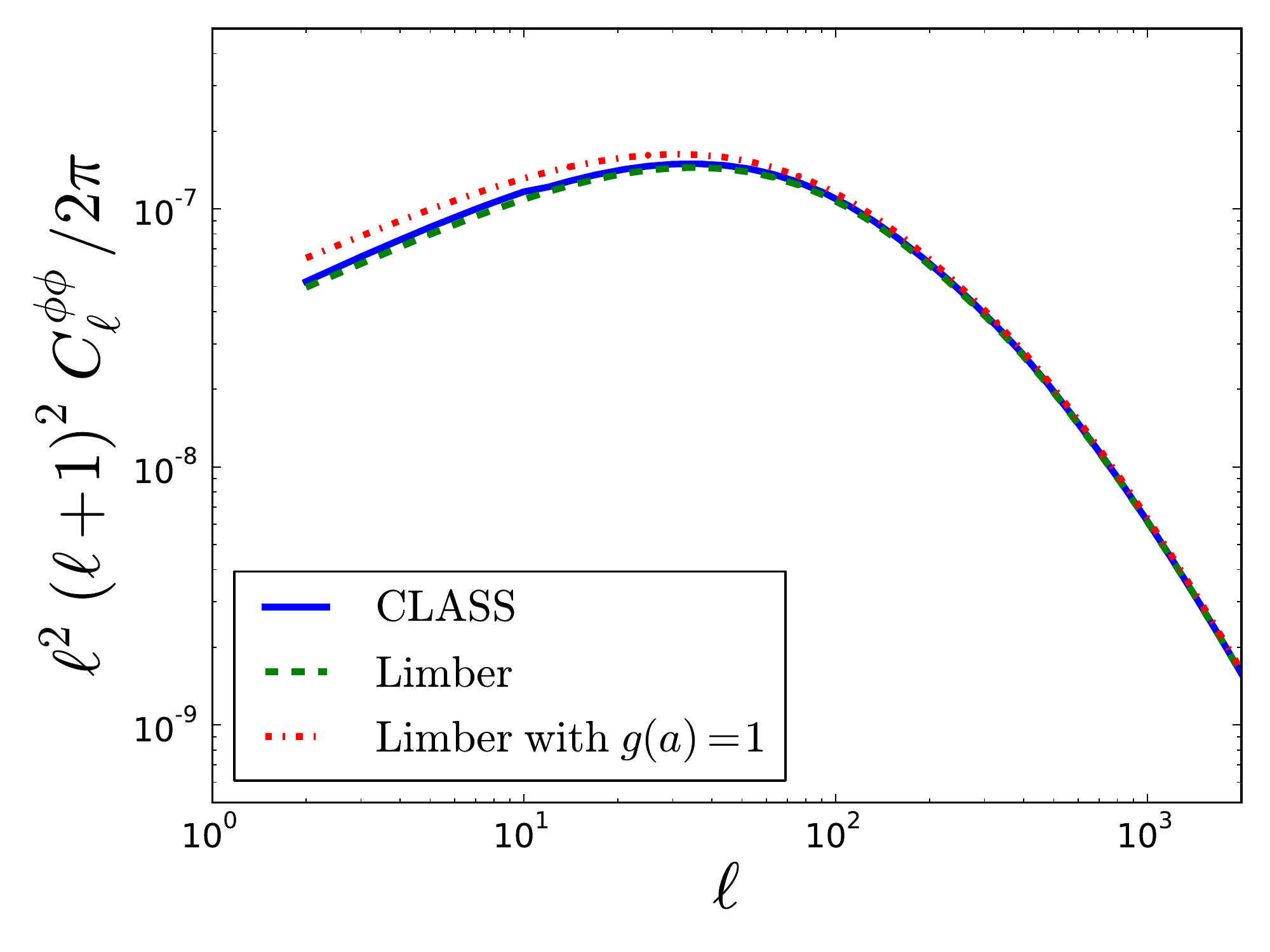}
\caption{The lensing power spectrum calculated from {\tt CLASS} (solid line),
calculated with Limber approximation (dashed line) and calculated with Limber approximation and setting $g(a)=1$ (dashed-dotted line).}
\label{fig:limbergood}
\end{figure}

With these pieces in place, we now examine the accuracy of the Limber approximation.
According to \citet{2008PhRvD..78l3506L}, it is a better approximation to
replace $k = \ell/\chi$ with $\sqrt{\ell(\ell+1)}/\chi\simeq(\ell+0.5)/\chi$ in the original Limber approximation Eq.(\ref{eqn:lensCl}). With the replacement and defining $x = \chi/\chi_\star$, the lensing power spectrum can be written as
\begin{\eq}
  \ell^2(\ell+1)^2 C_\ell^{\phi\phi} \simeq
  4\chi_\star\int_0^1 \mathrm{d}x\ (k^4P_\Phi)\left(\frac{\ell+0.5}{x\chi_\star};a\right)
  (1-x)^2 ,
\label{eqn:lensClx}
\end{\eq}
where $a=a(x)$.
Fig. \ref{fig:limbergood} shows the lensing power spectrum calculated from {\tt CLASS}
\citep{2011arXiv1104.2932L, 2011arXiv1104.2934L, 2011JCAP...07..034B, 2011JCAP...09..032L} compared to the lensing power spectrum calculated from the Limber approximation.
We see the Limber approximation reproduces the accurate numerical result even 
for small $\ell$.
In order to understand the the influence of the growth function, we also calculated
the lensing power spectrum by setting $g(a)\equiv1$. The growth function makes a 
difference for $\ell\la50$ \citep{2014MNRAS.445.2941P}. 
Now we are to understand the impact of massive neutrinos on $C_\ell^{\phi\phi}$.

\subsection{Influence of massive neutrinos on the lensing power spectrum: results}
\label{sec:lenmass}

According to Eq.(\ref{eqn:Pphi}) and Eq.(\ref{eqn:lensClx}), 
it is clear that $C_\ell^{\phi\phi}$ is determined by
the primordial perturbation $A_s, n_s$, the transfer function $T(k)$,
the impact of massive neutrinos $s(k; a)$, the growth factor $g(a)$,  
and the comoving distance to the last scattering surface $\chi_\star$.

\begin{figure}
\includegraphics[scale=0.42]{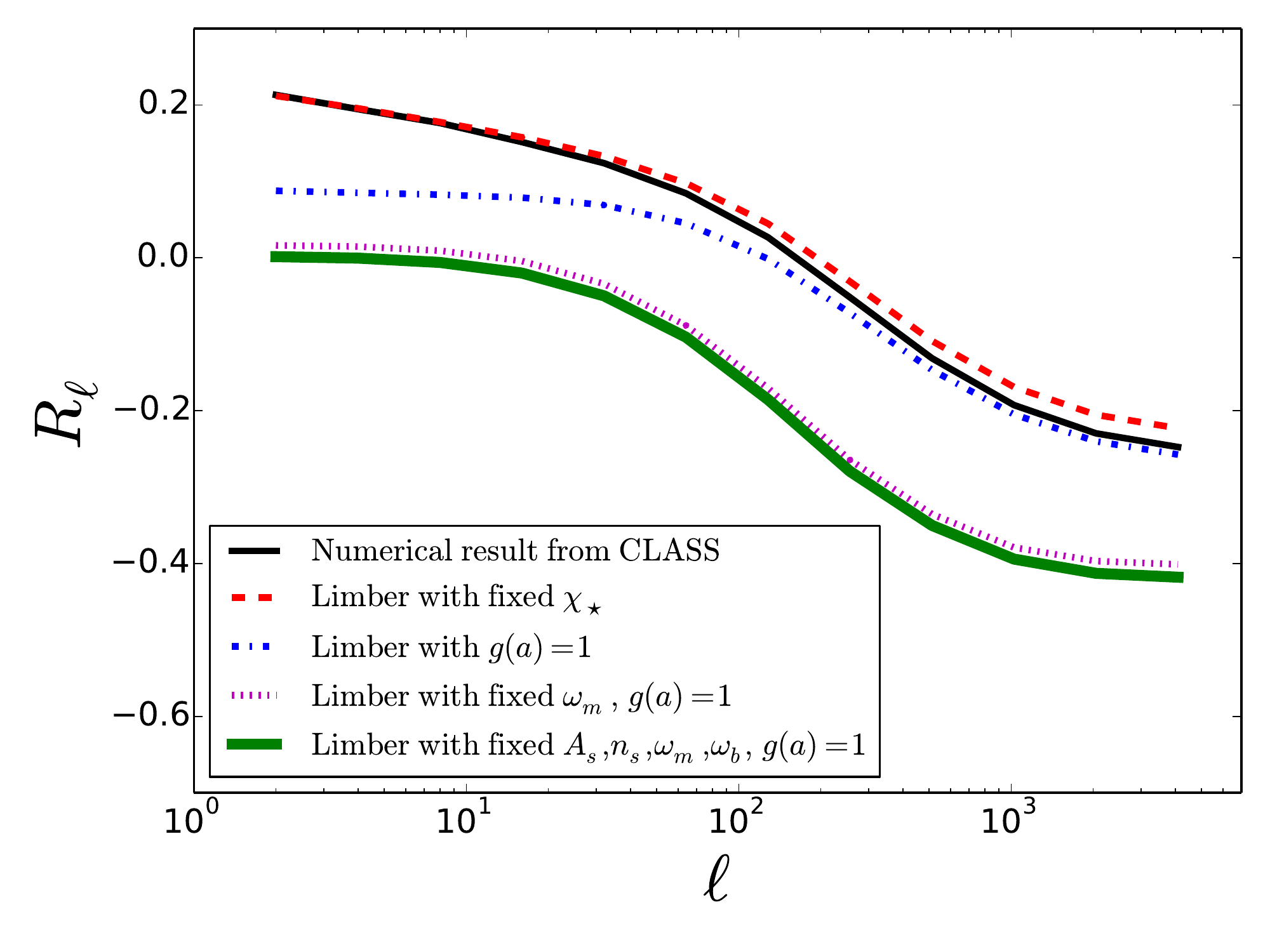}
\caption{The dependence of the lensing power spectrum on total neutrino mass, 
$(C_\ell^{\phi\phi}-C_{\ell, \rm fid}^{\phi\phi})/C_{\ell, \rm fid}^{\phi\phi}
= R_\ell \times (M_\nu/{\rm eV})$.
The black line is the numerical result from {\tt CLASS}, 
the red line is the result of Limber approximation setting $\chi_\star = 1.4\times 10^4 \ {\rm Mpc}$,
the blue line is the result of Limber approximation with $g(a)= 1$,
the magenta line is the result of Limber approximation fixing both $\omega_{\rm m}$ and $g(a)$,
and green line is the result of Limber approximation with $A_s, n_s$, $\omega_{\rm m}, \omega_b$ and $g(a)$ fixed.}
\label{fig:linear}
\end{figure}

In order to quantify the dependence of the lensing power on total neutrino mass $M_\nu$,
we take samples from a \Planck $\Lambda$CDM $+M_\nu$ chain, and fit the linear relation 
\be
\frac{C_\ell^{\phi\phi}-C_{\ell, \rm fid}^{\phi\phi}}{C_{\ell, \rm fid}^{\phi\phi}} = R_\ell \times M_\nu ({\rm eV}),
\ee 
The linear fitting result is shown as the thin solid line in Fig. \ref{fig:linear}.
To understand the contribution of the various parameter variations and effects, we also plot $R_\ell$ in Fig. \ref{fig:linear} for the cases shown.

We now work our way towards an understanding of the full response, the thin solid curve, in stages. We begin with the case where we fix most of the parameters other than $M_\nu$, and turn off the impact of dark energy on the growth factor, $g(a)$, fixing it to unity. In this case, increasing $M_\nu$ has two effects: 1) it decreases the free-streaming length $\sim (T_\nu/M_\nu)/H(z)$  \citep{2011APh....35..177A} , and 2) increases the expansion rate once the neutrinos start to become non-relativistic. The increased expansion rate acts to suppress the growth of structure on all length scales. However, the decreased free-streaming length acts to boost structure growth on scales above the free-streaming length, nearly exactly canceling the suppression. The result is the bottom-most curve of Fig.~\ref{fig:linear}: nearly no effect at low $\ell$, a constant suppression of power at high $\ell$, and a smooth transition between these two regimes.

The difference between the bottom-most curve and the dot-dashed curve (the one labeled, `Limber with  $g(a)=1$') is due to how other parameters adjust as $M_\nu$ varies. We can isolate these changes as almost all due (at least at $\ell \ga 50$) to a correlation between $M_\nu$ and $\omega_m$, as demonstrated by the following: if we fix $\omega_m$, and let $A_s$, $n_s$, and $\omega_b$ vary, we get the second curve from the bottom which differs very little from the bottom-most curve. Once we let $\omega_m$ vary as well, we get the dot-dashed curve. Letting $\omega_m$ vary leads to variation in $n_s$ and $A_s$ as well, but these changes all flow from the correlation between $\omega_m$ and $M_\nu$.

Once we let $g(a)$ vary as well we get a curve that is indistinguishable from the thin solid curve, which is boosted everywhere, and especially at $\ell \la 50$. The contributions to these large angular scales come predominantly from modes to the left of the peak in the matter power spectrum. At fixed (large) angular scale, structures that are nearer by, and therefore on smaller length scales, are closer to the peak of the matter power spectrum. Thus the large angular scales are weighted toward later times, and therefore more influenced by $g(a)$ than the smaller angular scales.  The growth factor {\em increases} with increasing $M_\nu$ because to keep the angular size of the sound horizon fixed $\Omega_\Lambda$ must decrease.

The lensing kernel's dependence on cosmological parameters comes entirely via its dependence on $\chi_*$.
To see how much of the variation in the lensing power spectrum is due to the lensing kernel, we fix $\chi_* = 1.4\times10^4$ Mpc (top-most dashed curve). By examining the difference between the top-most dashed curve, for which $\chi_*$ is fixed, and the thin solid curve, which is the full numerical result, one can see this effect is very small\footnote{This result is in contrast to the case of tomographic cosmic shear as a probe of dark energy. In this case the sensitivity of the data to variations in the dark energy equation-of-state parameter largely arises from the lensing kernel \citep{Simpson2005, Zhang2005}.

%Finally, performing the integral with the correct value of $\chi_*$ from each element of the chain results in a curve that is indistinguishable from the solid curve, except at low $\ell$ where the Limber approximation is not as accurate. The small difference between the top dashed curve and the solid curve shows that changes to the lensing kernel have very little to do with the sensitivity of the lensing power spectrum to variations in $M_\nu$.

To summarize, there are three main effects of massive neutrinos on the lensing power: 1) increased expansion rate suppresses power, 2) decreased free-streaming length compensates for the suppressed power at scales above the free-streaming length, 3) other parameter variations due to partial degeneracies in $C_l^{TT}$ (most notably an increase in $\omega_m$) boost the power on all scales. The net result is increased power at large scales and a decrease in power at small scales. One might potentially include the growth factor here as the fourth-most important effect, somewhat increasing the power at large angular scales.

The origin of the degeneracy in $C_l^{TT}$ between $\omega_m$ and $M_\nu$ is actually due to lensing itself. \citet{PlanckCollaborationXVI.2013} demonstrated that the dominant effect leading to the constraint of neutrino mass from the CMB temperature anisotropy power spectrum is gravitational lensing. As shown in Fig. \ref{fig:linear},  increasing $M_\nu$ suppresses the lensing power, while increasing  $\omega_{\rm m}$ increases the lensing power. The lensing power suppression by massive neutrinos can be compensated by the enhancement from increasing  $\omega_{\rm m}$, so uncertainties in $M_\nu$ and  $\omega_{\rm m}$ are expected to be positively correlated \citep{2010JCAP...12..027N}. }

\section{Forecast of constraints on the total neutrino mass from different data sets}
\label{sec:fisher}
We use the Fisher matrix formalism to forecast constraints on neutrino mass from future CMB and LSS experiments. The fiducial cosmology used here is the same as the one used in Section \ref{sec:galaxy} except with a different value of total neutrino mass, $M_\nu=85$ meV. 

\subsection{CMB-S4 and DESI BAO}
Following \citet{2014ApJ...788..138W} and \citet{Dodelson03}, 
the Fisher matrix for cosmological parameters constrained by CMB spectra is written as
\bea
F_{\alpha\beta} = \sum_\ell^{\ell_{\rm max}} \frac{2\ell+1}{2} f_{\rm sky}{\rm Tr}
\left(\C^{-1} \frac{\partial \C} {\partial \theta_\alpha}\C^{-1} \frac{\partial \C} {\partial \theta_\beta} \right),
\eea
and it is related to the expected uncertainty of a parameter $\theta_\alpha$ by $\sigma(\theta_\alpha) = \sqrt{(F^{-1})_{\alpha\alpha}}$,
where 
\be
\C = \left( \begin{array}{ccc}
C_\ell^{TT}+ \N^{TT} & C_\ell^{TE} & C_\ell^{Td}\\
C_\ell^{TE} & C_\ell^{EE}+ \N^{EE} & 0\\
C_\ell^{Td} & 0 & C_\ell^{dd}+ \N^{dd} \end{array}\right),
\ee
and $C_\ell^{dd}$ is the angular power spectrum of the deflection field $\mathbf{d}$,
which is related to the lensing power spectrum by 
$C_\ell^{dd} =\ell(\ell+1)C_\ell^{\phi\phi}$. The Gaussian noise $\N^{XX}$ is defined as
\be
\N^{XX} = \Delta_X^2 \exp \left(\ell(\ell+1)\frac{\theta^2_{\rm FWHM}}{8\log 2}\right),
\ee
where $\Delta_X$ ($X= T, E, B$) is the pixel noise level of the experiment 
and $\theta_{\rm FWHM}$
is the full-width-half-maximum beam size in radians \citep{Knox1995,1997ApJ...488....1Z}.
The noise power spectrum of deflection field  $\N^{dd}$ is calculated 
assuming a lensing reconstruction that uses the quadratic $EB$ estimator\citep{2003PhRvD..67h3002O}. 
We use the iterative method proposed by \citet{2012JCAP...06..014S},
which performs significantly better than the uniterated quadratic estimators \citep{2003PhRvD..68h3002H}.

\begin{figure}
\includegraphics[scale=0.4]{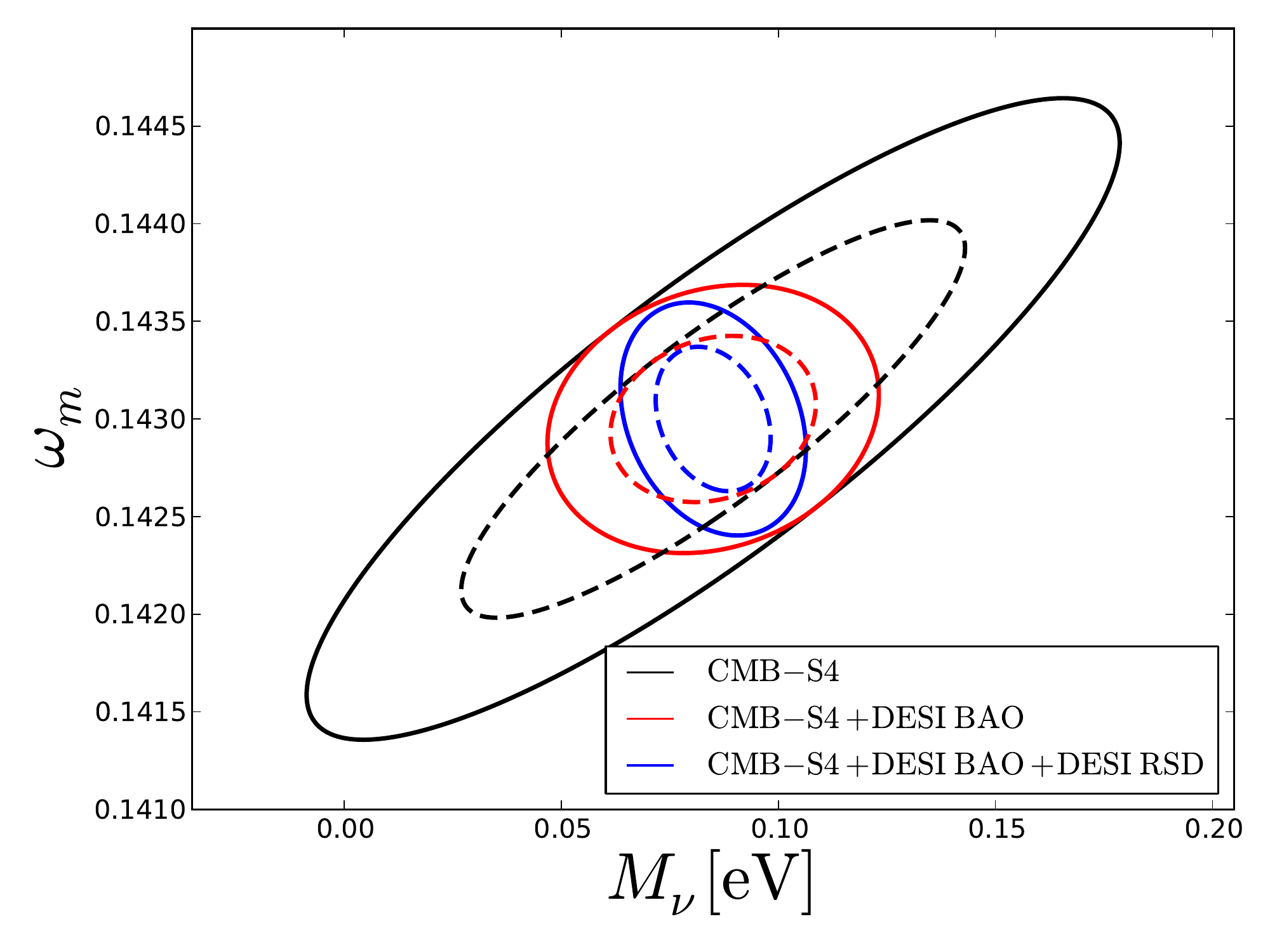}
\caption{Forecasted $1\sigma$ and $2\sigma$ constraints in the $M_\nu-\omega_{\rm m}$ plane,
where the CMB-S4 experiment results in a $\sigma(M_\nu) = 38$ meV constraint,
the combination of CMB-S4 and DESI BAO yield a $\sigma(M_\nu) = 15$ meV constraint.
and adding measurements of the structure growth rate by DESI RSD further improves the constraint to $\sigma(M_\nu) = 9$ meV. }
\label{fig:constraint}
\end{figure}

For the CMB-S4 experiment, we assume the temperature noise level
$\Delta_T = 1.5 ~\mu$K-arcmin, the polarization noise level 
$\Delta_E =\Delta_B = \sqrt{2} ~\Delta_T$, the fraction of covered sky 
$f_{\rm sky} =0.5$ and the beam size $\theta_{\rm FWHM}=1'$.
With these given experiment sensitivities, we obtain a constraint from 
CMB with $\sigma(M_\nu) = 38$ meV. The $1 \sigma$ and $2 \sigma$
constraint are shown in Fig. \ref{fig:constraint}.

According to the analysis in Section \ref{sec:galaxy} and \ref{sec:lensCl} , DESI BAO  are helpful to break the degeneracy between $M_\nu$ and $\omega_{\rm m}$. BAO uncertainties are independent from CMB experiments, so the total Fisher matrix is simply given by addition
\be
F_{\rm CMB+BAO} = F_{\rm CMB} + F_{\rm BAO},
\ee
where the DESI sensitivities of BAO signal can be found in \cite{2014JCAP...05..023F} and shown in Fig. \ref{fig:DzHz}.
It is found that, adding the DESI BAO data greatly improves the constraint to $\sigma(M_\nu) = 15 $ meV (similar forecasts were also conducted by \cite{2013arXiv1309.5383A, 2014ApJ...788..138W}).

\subsection{Beyond DESI BAO}
The largest signal of massive neutrinos on $H(z)$ and $D(z)$ is found at low redshifts (see Fig. \ref{fig:DzHz}), where BAO has inevitably large noise because of small amount of survey volume and large cosmic variance. Other than DESI BAO, we also investigate other low-redshift tracers of $H(z)$ and $D(z)$ which are possible to tighten the uncertainty of total neutrino mass. 

DESI RSD: similar to BAO, RSD uncertainties are also independent from those of CMB observations, so the total Fisher matrix of  CMB+BAO+RSD is also approximately given by addition 
\be
F_{\rm CMB+BAO+RSD} = F_{\rm CMB} + F_{\rm BAO}+ F_{\rm RSD},
\ee
where we use the RSD sensitivities from DESI survey which can be found in \cite{2013arXiv1309.5385H} and shown in Fig. \ref{fig:growth}.
Here we use the approximation that uncertainties in BAO and RSD are uncorrelated, due to they are sensitive to different aspects of the matter power spectrum: BAO is sensitive to its characteristic length scale $r_s$ while RSD is sensitive to its amplitude. In fact, our result is insensitive to the approximation because we find that both CMB-S4+DESI BAO+DESI RSD and CMB-S4+DESI RSD yield the same $\sigma(M_\nu)=9$ meV uncertainty. 

\begin{figure}
\includegraphics[scale=0.4]{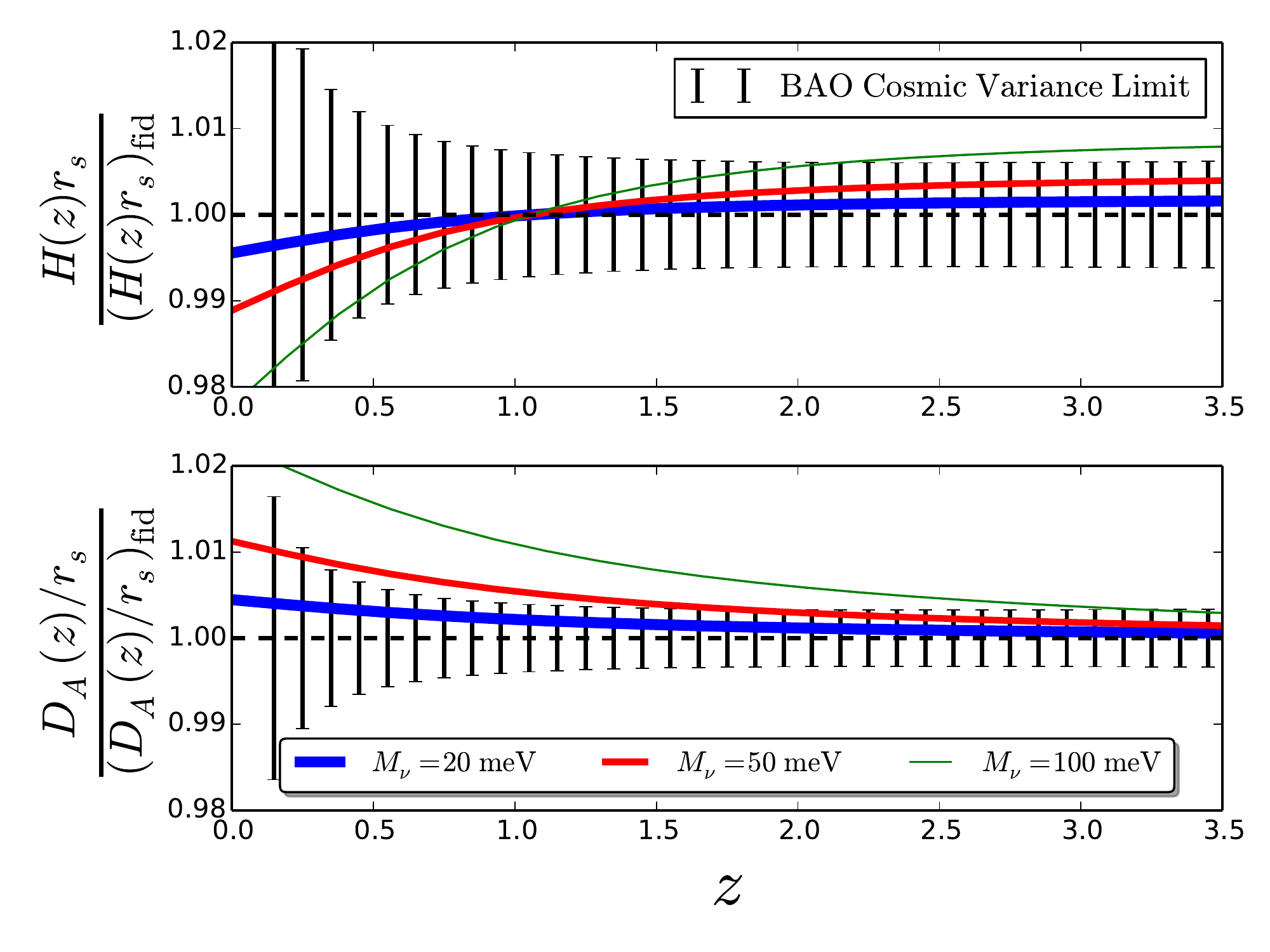}
\caption{Same as Fig. \ref{fig:DzHz}, but with suppressed errorbars of $D_A(z)$ and $H(z)$ coming from CMB-S4 and a cosmic-variance-limited BAO experiment. }
\label{fig:bestbao}
\end{figure}

\emph{Better} BAO: DESI survey cover $14,000$ squared degrees (about $1/3$ of the whole sky). We explore a future BAO experiment which covers the whole sky and in which cosmic variance dominates over shot noise in the redshift range $0 < z < 4.0$.  Constraints on $D_A(z)$ and $H(z)$ from this BAO experiment are  shown in Fig. \ref{fig:bestbao}. It is found that CMB-S4 and the cosmic variance limited BAO constrain the total neutrino mass with uncertainty $\sigma(M_\nu) = 11$ meV. So we conclude that $11$ meV is a lower limit of $\sigma(M_\nu)$ we could measure from CMB-S4+BAO, where the limit mainly comes from noise level of the CMB lensing signal.
 
\begin{figure}
\includegraphics[scale=0.4]{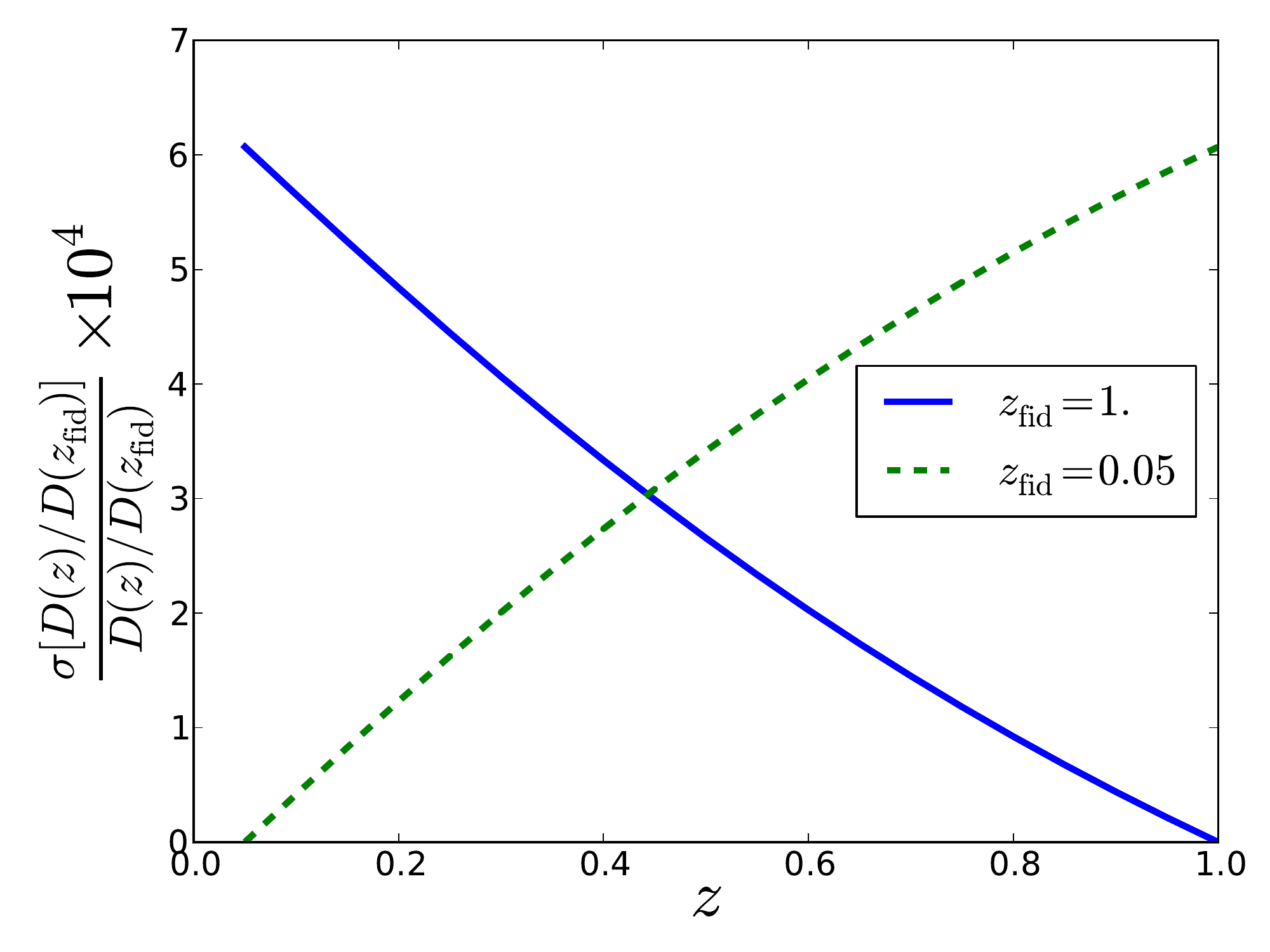}
\caption{The uncertainties in relative distances from CMB-S4 + DESI BAO. Note that the uncertainties is multiplied by a factor of $10^4$ in the plot.}
\label{fig:sn}
\end{figure}

\emph {Supernovae}: The constraining power of BAO is limited by its large cosmic variance at low redshifts (Fig. \ref{fig:bestbao}), so supernovae distance measurements which do not suffer from the cosmic variance problem may be effective complements if their systematic errors are well controlled. Supernovae perform better in relative distance measurements than in absolute distance measurements. However for the $\Lambda$CDM + $M_\nu$ model, the uncertainties in relative distances from CMB-S4 + DESI BAO are very small (see Fig. \ref{fig:sn}) . We conclude that supernova observations must result in relative distance determinations with
systematic errors less than about  $0.05\%$ if they are to
tighten the constraints on neutrino mass. Compared to 
systematic errors from current supernova observations
(e.g., \citet{Suzuki2012}) this would be a reduction by 
a factor of $\sim 20$ . 

\section{Conclusion}
This paper is motivated by our desire to better understand the origin of current and forecasted cosmological constraints on the sum of neutrino masses. We took as a given that determination of $N_{\rm eff}$ will solidify the predicted value of 3.046, increasing our confidence that the phase-space distribution of the cosmic neutrino background is what we expect based on the standard thermal history. With that as a given, the most important aspect of increased neutrino mass (relative to some reference model) is an increased neutrino energy density. If the model with increased mass is to remain consistent with CMB observations, the distance to last-scattering must be preserved and so the total energy density, and therefore the expansion rate, cannot increase at all redshifts. To compensate for the increase in neutrino energy density, the cosmological constant must decrease in value. Thus varying neutrino mass leads to changes in $H(z)$ with  a very particular shape: a mild increase at high redshifts, a larger decrease in low redshifts, with a transition near the onset of $\Lambda-$domination at $z \simeq 1$. Unfortunately this very specific prediction of the shape for $H(z)$ is difficult to verify in detail because of how small the departures from the reference model are at $z > 1$. We see that this difficulty persists even for a cosmic-variance limited all-sky ($z < 4$) BAO experiment (see Fig. \ref{fig:bestbao}).

The sensitivity of CMB-S4 to neutrino mass comes via the impact of this increased expansion rate on the growth of structure.
At scales below the neutrino free-streaming length, this increased expansion rate suppresses the growth of structure. Above the
free-streaming length, the ability of massive neutrinos to cluster compensates for the increased expansion and there is no net suppression.
Because increasing matter density  increases lensing power amplitude, the CMB lensing-derived constraints on neutrino mass have uncertainties positively correlated with the matter density uncertainties. This correlation with dark matter density leads to secondary
correlations of neutrino mass uncertainty with uncertainties in $n_s$ and $A_s$. We disentangled all these various effects in
Fig.~\ref{fig:linear}.

The correlation between $M_\nu$ and $\omega_{\rm m}$ has the opposite sign as that from BAO, since increasing neutrino mass and increasing $\omega_{\rm m}$ both increase the expansion rate at $z > 1$, and lead to a compensating decrease at $z < 1$. Thus the combination of CMB-S4 and DESI-BAO leads to improvements in the determination of both quantities.

Finally, we briefly investigated how constraints might be improved beyond the  $\sim 3\sigma$ to  $\sim 4\sigma$ detection expected from CMB-S4 + DESI BAO in the case of the lowest possible neutrino mass of $58$ meV. The larger signals are at low redshift, so we considered supernovae as relative distance indicators and redshift-space distortions, as well as cosmic-variance-limited BAO. We found requirements on supernova precision that are probably prohibitively stringent. But better BAO, and RSD,  both have the potential to improve the detection to $\sim 5 \sigma$ or greater.

\section{Acknowledgments}
We thank Anze Slosar and Patrick McDonald for sharing the forecasts from DESI BAO.
Z.P. would like to thank Brent Follin for introducing {\tt CLASS} code and Marius Millea for python package {\tt Cosmoslik}. This work made extensive use of the NASA Astrophysics Data System and of the {\tt astro-ph} preprint archive at {\tt arXiv.org}.

\bibliographystyle{mn2e} 
\bibliography{ms}

\end{document}